\documentclass[reprint,amsmath,amssymb,aps]{revtex4-2}

\usepackage{graphicx}%
\usepackage{dcolumn}%
\usepackage{bm}%
\usepackage{float}
\usepackage{amsmath}

\usepackage{xcolor}

\begin{document}

\title{Inverse design of self-folding 3D shells}

\author{Diogo E. P. Pinto$^{1}$, Nuno A. M. Ara\'{u}jo$^{2,3}$, Petr \v{S}ulc$^{4,5}$, John Russo$^{1}$}

\affiliation{$^{1}$ Dipartimento di Fisica, Sapienza Universit\`{a} di Roma, P.le Aldo Moro 5, 00185 Rome, Italy \\ $^{2}$ Centro de Física Teórica e Computacional, Faculdade de Ciências, Universidade de Lisboa, 1749-016 Lisboa, Portugal. \\ $^{3}$ Departamento de Física, Faculdade de Ciências, Universidade de Lisboa, 1749-016 Lisboa, Portugal.\\ $^{4}$ School of Molecular Sciences and Center for Molecular Design and Biomimetics, The Biodesign Institute, Arizona State University, 1001 South McAllister Avenue, Tempe, Arizona 85281, USA\\  $^{5}$ TU Munich, School of Natural Sciences, Department of Bioscience, Garching, Germany}

\begin{abstract}

Self-folding is an emerging paradigm for the inverse design of three-dimensional structures. While most efforts have concentrated on the shape of the net, our approach introduces a new design dimension—bond specificity between the edges. We transform this design process into a Boolean Satisfiability problem to derive solutions for various target structures. This method significantly enhances the yield of the folding process. Furthermore, by linearly combining independent solutions, we achieve designs for shape-shifting nets wherein the dominant structure evolves with varying external conditions. This approach is demonstrated through coarse-grained simulations on two examples of triangular and square nets capable of folding into multiple target shapes.

\end{abstract}

\maketitle

The ability to externally control the formation of microscopic structures and to selectively switch between different conformations are among of the most ambitious goals of materials design~ \cite{Whitelam2015, Meng2010, Garmann2022}. One of the most successful paradigms for the bottom-up realization of ordered aggregates, from the molecular to the colloidal scale, is self-assembly~\cite{Pandey2011, Frenkel2011, Sartori2020}. In this process, a dispersion of building blocks aggregate due to carefully designed attractive (or entropic~\cite{vo2022theory}) interactions.
Despite its potential, self-assembly is inherently limited by the kinetics of the aggregation process, and is often derailed by the presence of kinetic intermediate structures whose long lifetime prevents the correct assembly of the target structure~\cite{Dodd2018, Joshi2016, Bupathy2022}.

\begin{figure*}[t]
	\includegraphics{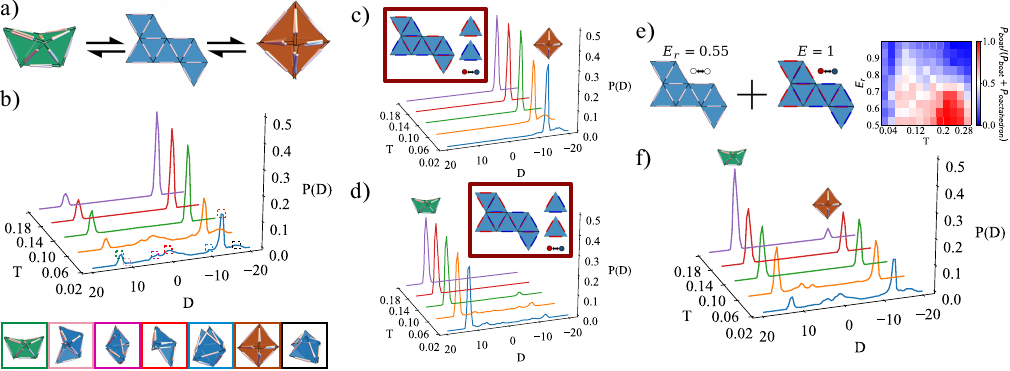}
	\caption{\label{fig1} a) Schematic representation of the triangular net used, with the two possible shells that can be folded. On the left is the octahedron while on the right is the boat. The rods on the edges are only used for visualization purposes. b) Histogram of the order parameter for different temperatures. Below the histogram are represented the most probable misfoldings with a color scheme corresponding to the points highlighted in the histogram. c) Histogram of the order parameter as a function of temperature for a design which excludes the boat. d) Similar to c) but with a design which excludes the octahedron. The respective designs are highlighted by the red box, which includes the net, the different tiles used and the interaction matrix between the colors. e) Schematic showing the linear combination of two nets that compete to fold the shells. On the right is shown the fraction of boats formed between all closed structures for different temperatures and $E_r$. f) Histogram of the order parameter as a function of temperature for the linear combination of nets shown in d). All histograms are averages over 1000 different samples, while the colormap uses 100 different samples.}
\end{figure*}

Inspired by the tremendous progress of DNA nanotechnology in general and DNA origami in particular, here we investigate an alternative route to self-assembly, represented by \emph{self-folding} materials, i.e. 2D planar templates (\emph{nets}) that are designed to fold into the desired structure~\cite{Faber2018, kim2023harnessing}. Compared to traditional self-assembly methods, self-folding has some key advantages: i) it  can be triggered on faster time-scales compared to self-assembly, as the different units do not need to explore the volume of the system to assemble; ii) its basic units, the \emph{tiles}, are generally simpler to realize, as one needs only to consider planar interactions, compared to the complex three-dimensional building blocks required for self-assembly; iii) it holds the promise to realize \emph{shape-shifting} materials, as self-folding can easily adapt to changing external conditions, contrary to the structures obtained by self-assembly which are difficult to reconfigure without disassembling and reassembling the components.

Experimentally, DNA-origami have made the biggest contribution towards fully realizing the potential of self-folding systems thanks to their nanoscale precision, geometric design, and the fully-controllable specificity of the base pairing. Moreover, they allow for selective interactions between elementary constituents through different energy scales, by tuning DNA chain length~\cite{Geerts2010}. Application to produce controllable planar nets have seen a rapid growth in recent years, and include so called DNA kirigami~\cite{chen2022dna}, DNA origami tesselations~\cite{Yue2023}, and recently introduced reconfigurable (akin to paper-folding mechanism) planar DNA origami~\cite{kim2023harnessing}, to name a few. Additionally, new micron-sized planar structures based on seeded assembly of DNA origami criss-cross slats \cite{Wintersinger2023} would present a new potential way to create larger shape- shifting 3D nanostructures folded from 2D planar template. 
The folding of 2D planar templates has also gathered a lot of theoretical interest, especially as a means of assembling 3D capsules or shells~\cite{Sussman2015}. 
Given that the phase space of structures that can be folded is finite and well-known \cite{Araujo2018} it is the folding pathway that dictates the final structure.
Control over the folding pathways for single~\cite{Melo2020, Simoes2021} and multiple targets~\cite{Azam2009} have so far focused on the influence of the network topology on the final assembled structure~\cite{Dodd2018}.

In this work we propose a novel way to control the folding pathway of 2D planar nets based on bond specificity between the edges of the tiles that compose the net. Our goal is to enhance and/or selectively steer the folding process by optimizing the interaction between the edges for any given net.
To demonstrate our approach we first focus on a prototypical example of self-folding net, shown in Fig.~\ref{fig1}a, which is composed of eight faces, all being equilateral triangles of the same size. The net can fold into two ordered 3D shells, the octahedron and the boat, and a large variety of disordered structures.
Each triangular face of the net represents a tile, where edges between tiles can interact attractively. The interactions can be represented as \emph{colors}, such that two edges interact according to a color interaction matrix. 
This coloring is subject to multiple constraints. Firstly, we require the nets to be able to tile the plane, so that one can first create a 2D triangular lattice with the respective tiles and then cut the final net to fold the 3D shell; this condition mimics standard experimental methods where a plane is first seeded on a 2D substrate and then the final net is etched from it \cite{Chen2020}. Secondly, we want the net to fold in either the boat or octahedron configurations~\cite{Meng2010, Azam2009}. Alternatively, we want to be able to change the target  structure depending on the external conditions, i.e. create reconfigurable shape-shifting structures. 

To tackle the computational complexity imposed by these conditions,
we use an inverse design method called SAT-assembly \cite{Russo2022}, which translates the topologies of the 3D target shells into a Boolean Satisfiability Problem (SAT), such that the design problem is formulated in terms of binary variables and logic clauses, for which fast solution methods are available \cite{Een2005} .
In particular, we consider that each edge can be attributed a color $x_c \in\{1,2, ...,N_c\}$, where $N_c$ is the total number of distinct colors used among all tiles. The tile type $x_p$ is specified by the color arrangement of its edges, with each unique combination representing a different type $x_p \in\{1,2, ...,N_p\}$, where $N_p$ is the total number of different tile types. 
$N_c$ and $N_p$ are the input parameters. We then use a SAT solver \cite{Een2005} to find the tile coloring with $N_c$ colors and $N_p$ types that satisfies all the constraints.
In the following, we will show results for the system of Fig.~\ref{fig1}. In the \emph{Supplementary Material} we go into more detail on the constraints (clauses) used in SAT.

To validate the folding pathways, we conducted molecular dynamics simulations of the self-folding process. To model the tiles, we employed a patchy particle model~\cite{russo2022physics}, where each face is represented by a hard-core sphere with a radius of $0.3$ (in units of length corresponding to the distance from the center of each tile to any of its vertices, denoted as $\sigma$). This choice aims to minimize steric effects while preventing face overlap. Two attractive patches are positioned on each edge at equidistant points from the center of the edge. This minimum number of patches per edge is necessary to enable face hinging. The patches exhibit an attractive isotropic point potential to characterize the attractive interactions between the edges. If two patches, of the same color, are within a range of $\delta=0.18$, they form a bond with an energy of $\varepsilon$ (our energy unit). The internal edges of the net (those initially bonded) interact with an energy of $5,\varepsilon$ to ensure that they do not break across the temperature range explored in this study.
More details on the potentials can be found in \textit{Supplementary Material}.
We perform Brownian Dynamics simulations of our model using the oxDNA package \cite{oxDNA}. The particle system was simulated using rigid-body molecular dynamics with an Andersen-like thermostat \cite{Russo2009}. Temperature is measured in units of $\epsilon/k_B$, where $k_B$ is the Boltzmann constant. We always start from the same initial configuration which corresponds to a flat net (Fig.~\ref{fig1}a) and run the simulations for a maximum of $10^9$ timesteps, with each step corresponding to $\Delta t=0.001$, in units of $\sigma / \sqrt{m/\varepsilon}$, where $m$ is the mass of each individual tile. During the simulation, each patch on the edge was only able to bind to one other at a time. All results are averaged over multiple independent simulations.

\begin{figure}[t]
	\includegraphics{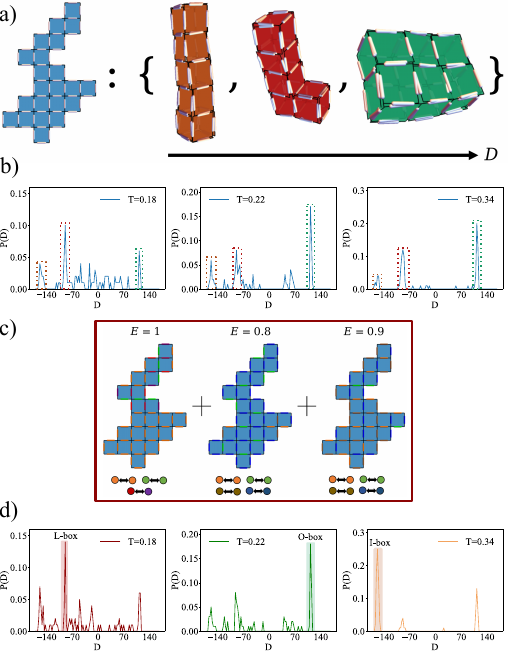}
	\caption{\label{fig2} a) Schematic representation of the square tile net which can fold three different structures, the \textit{I-box}, the \textit{L-box} and the \textit{O-box}, which are ordered according to their order parameter value. In b) we show histograms of the order parameter for different temperatures for the N1c1 design and highlight with colored boxes the corresponding peaks for the folded shells. c) Schematic of the linear combination of three nets with the respective interaction energies and the interaction matrix between the colors. d) Histograms of the order parameter for different temperatures of the linear combination design proposed. Each line is colored according to the most probable folded shell. Results were averaged over 100 different samples.}
\end{figure}

We first consider a net where all edges have the same color that binds to itself, as shown in Fig.~\ref{fig1}b. We refer to this design as N1C1 (one tile type and one color). Given that we always start with the same net, there is only one possible combination for bonding between patch pairs ($pp$) that closes either the octahedron ($O$) or the boat ($B$). We can create the contact network for either structure, which includes the respective patch pairs, and check, after folding, if the shell formed satisfies either one structure or the other. To properly identify the different shells, we introduce an order parameter based on the distance between pairs of patches:

\begin{equation}
\label{eq1}
    D=D_{\rm octahedron}-D_{\rm boat}=\sum_{pp \in O} r_{ij} - \sum_{pp \in B} r_{ij} \ ,
\end{equation}

\noindent where $D_{\rm octahedron}$ is equal to the sum of the distances ($r_{ij}$) between all patch pairs in the octahedron contact network, while $D_{\rm boat}$ is equal to the sum of the distances between all patch pairs in the boat contact network. Thus, if a net folds a boat, $D_{\rm boat}\ll D_{\rm octahedron}$ and $D$ is largely positive, while if it folds an octahedron, $D_{\rm octahedron}\ll D_{\rm boat}$ and $D$ is largely negative. The order parameter
also gives information if a given misfolded shell is closer to the octahedron or the boat. In Fig.~\ref{fig1}b we show a histogram of the order parameter for different temperatures.
In the snapshots we highlight the octahedron, the boat, and the most probable misfolded shells observed as peaks in the order parameter distribution. We observe that at low temperatures the system often gets trapped in misfolded configurations. As temperature is increased, thermal fluctuations allow the system to find the two free energy minima corresponding to the boat and octahedron configurations. From the height of the peaks, we observe that the octahedron is the kinetically preferred structure for this model, even if both the octahedron and boat have the same number of bonds and, thus, the same energy.
The results of Fig.~\ref{fig1}b highlight the biggest problem with generic self-folding systems: the presence of multiple local minima that can significantly reduce the yield of the final structure, especially at low temperatures, where the aggregates are kinetically stabilized.

In Fig.~\ref{fig1}c and d we show the effect of using colored designs, using SAT to find a minimum combination of tile types that satisfies only one of the structures while avoiding the other. For both structures, we find that the minimum design requires two different tile types and two colors (N2c2). In the red box of Fig.~\ref{fig1}c is a design that folds the octahedron while avoiding the boat. The interaction matrix between the colored edges is also shown. Particularly, if an edge with the color red finds another with the color blue they interact as before, while if both edges have the same color the interaction energy is zero. In the histograms, we confirm that the probability of finding the boat is always zero.
In the red box of Fig.~\ref{fig1}d is a design that folds the boat while excluding the octahedron, as confirmed by the order parameter histograms.
The histograms show that this coloring strategy is not only effective at selecting the intended designs but also at suppressing disordered kinetic aggregates. As the specificity of the interactions increases, it is less probable that a bond not present in the final structure is formed.
Increasing the total number of colors/tile types is thus a viable strategy to increase the success probability of the folding process.

So far, we have used designs that target a specific structure while avoiding competing ones \cite{Bohlin2023, Pinto2023}. We now show how to use the designs produced by SAT to target multiple structures.
For that, it is sufficient to linearly combine the interaction matrices of two designs that target different structures. The coefficients of this linear combination express the relative strength of the bonds in one of the structures compared to the others.
For example, we can use two designs, the N1C1 (Fig.~\ref{fig1}b) and the N2C2 (Fig.~\ref{fig1}d), and assign their respective interaction matrices different energy values. Since the octahedron is more probable in the N1C1 design, we assign a higher energy value to the N2C2 design which only forms the boat. This means that the minima corresponding to the boat will be deeper and thus more energetically favored, while the dynamical pathways will favor the octahedron, as noted previously. With this method, if two edges with interacting colors according to the N2C2 design meet, they form a bond with energy $\varepsilon=1$. If they do not interact according to the N2C2 design, then they will interact only through the N1C1 design and have energy $E_r\leq1$.

In Fig.~\ref{fig1}e we show the two nets used with their respective coloring and the respective energies. We also show the order parameter histogram for different temperatures in Fig.~\ref{fig1}f. Here, we compare the folding of the boat and octahedron for $E_r=0.55$ (see \emph{Supplementary Material} for other results). We observe that, at lower values of temperature, the results are quite similar to the N1C1 shown in Fig.~\ref{fig1}b, where the octahedron is the most probable structure. Given that the temperature is lower, the folding process can get stuck in local minima for longer. Since the octahedron pathway is more probable than the boat, this shell will fold more frequently and due to the low temperatures remain stuck in this state. At higher values of temperature, the boat eventually becomes the most probable structure. Since the bonds break more frequently, the net will more quickly reach the global energy minimum (boat), which becomes more and more favored as $E_r$ decreases. Thus, we find that by introducing the two competing designs with relative energies one can favor different shells at opposite temperature ranges. In Fig.~\ref{fig1}e we also show a diagram of the parameter space for different $E_r$ and temperatures. The colormap indicates the fraction of boats formed and is calculated using $P_{boat}/(P_{boat}+P_{octahedron})$, where $P_x$ corresponds to the probability of folding a given structure.
The diagram shows that the free energy minimum associated with the boat configuration becomes dominant for
$E_r\leq0.7$, where the shape-shift between the octahedron and the boat can be controlled by varying the temperature. It is important to note that by quenching the system one cannot reverse an already folded shell, thus the net should be opened first by increasing the temperature. In the Supplementary Material we show results for temperature cycling of this net to control the most probable structure.

In Fig.~\ref{fig2} we show that this concept of linear combination of colored designs can be generalized to nets that can fold in more than two complete structures. In particular we show the results for the net shown in Fig.~\ref{fig2}a which is formed by square tiles and can fold three different shells~\cite{xu2017common}: two cuboids, the \textit{I-box} and the \textit{O-box}~\cite{nets}, and an L-shaped box, \textit{L-box}. We use a similar order parameter to the one introduced in Eq.~\ref{eq1}, $D=D_{I-box}-D_{O-box}$. Since the L-box is a combination of the other two shells, the peak corresponding to it will be located between the other two. In Fig.~\ref{fig2}b we plot the histograms for different temperatures corresponding to the folding of this net with the N1c1 design. At low temperatures the \textit{L-box} is favored while the \textit{I-box} is always suppressed by the others. In Fig.~\ref{fig2}c we show three designs which we linearly combine to control the folding process. The red box in the figure contains the colored nets, which are made of two different tiles and four colors (N2c4), and the corresponding color interaction matrices that respectively allow the assembly of the \textit{I-box}, \textit{L-box} and \textit{O-box}.
Utilizing SAT allows us to select linearly independent colored nets, implying that each shell cannot be derived by linearly combining solutions from the other shells. The Supplementary Material provides an explanation of the SAT algorithm for this procedure. Since the coefficient of the linear combination (representing the energy) specifically weights one particular structure, this property facilitates the temperature-based selection of the most probable shell, as depicted in the histogram in Fig.~\ref{fig2}d. At low temperatures the \textit{L-box} is the most probable, but at intermediate ones it changes to the \textit{O-box}, while at high temperatures the \textit{I-box} becomes more probable, showing the shape-shifting capability of the design. Given that the selected colored designs are linearly independent, it is possible to set one of the interaction energies to zero and only form two of the shells.

To conclude, we have shown how specific (colored) interactions can greatly improve the self-folding yield, and how several important properties can be embedded into the design via satisfiability methods. These allow us to suppress unwanted structures at the expense of a minimal increase in complexity.
We have then introduced the idea of linearly combining independent designs to achieve external control over the final structure. In our case, we can change the ratio between the different complete structures by changing the temperature.
With recent experimental advancements in DNA-origami nanotechnology, self-folding holds the potential to find wide application in the assembly of reconfigurable systems.
A foldable 2D template can be achieved from a single DNA origami \cite{kim2023harnessing}, or from multiple DNA origami nanostructures connected \cite{Wintersinger2023,tikhomirov2017fractal}. The edges can then be functionalized with single-stranded DNA overhangs that will act as bonds between edges, with specificity and interaction strength given by the interaction matrix from our SAT-assembly approach.
In this context, the SAT-assembly approach is a promising step towards programming variable shapes directly into the structure. It is possible to use our method to store multiple structures in a single 2D net thus enabling potential experimental realization of shape-shifting nanostructures that have multiple stable folds that can be selectively recalled \cite{murugan2015multifarious,fink2001many}. They can be designed to switch conformation at e.g. at different temperatures, or due to an external stimulus, such as the presence of single-stranded DNA detectors that can strengthen the interaction between specific edges that will drive the rest of the structure to refold \cite{kim2023harnessing,lowensohn2020self}.

\subsection*{Acknowledgements}

DEPP and JR acknowledge all the financial support from the European Research Council Grant DLV-759187. J.R. acknowledges support from ICSC — Centro Nazionale di Ricerca in High Performance Computing, Big Data and Quantum Computing, funded by the European Union—NextGenerationEU. NA acknowledges financial support from the Portuguese Foundation for Science and Technology (FCT) under Contracts no. UIDB/00618/2020 and UIDP/00618/2020. P\v{S} acknowledges funding from the European Research Council (ERC) under the European Union’s Horizon 2020 research and innovation programme (Grant agreement No. 101040035). 

\bibliography{refs}

\end{document}